\documentstyle[aps,preprint]{revtex}
\begin{document}
\draft \tightenlines
\preprint{gr-qc/0003047}
\draft\tightenlines

\title{Black Hole Decay and Quantum Instantons}
\author{Dongsu
Bak,$^{1}$\footnote{Electronic address: dsbak@mach.uos.ac.kr} Sang
Pyo Kim,$^{2}$\footnote{Electronic address:
sangkim@ks.kunsan.ac.kr}, Sung Ku Kim,$^{3}$\footnote{Electronic
address: skkim@mm.ewha.ac.kr} Kwang-Sup
Soh,$^{4}$\footnote{Electronic address: kssoh@phya.snu.ac.kr} and
Jae Hyung Yee,$^{5}$\footnote{Electronic address:
jhyee@phya.yonsei.ac.kr}}
\address{a Department of Physics, University of Seoul, Seoul
130-743 Korea\\ b Department of Physics, Kunsan National
University, Kunsan 573-701 Korea\\ c Department of Physics, Ewha
Women's University, Seoul 120-750 Korea\\ d Department of Physics
Education, Seoul National University, Seoul 151-742 Korea\\ e
Department of Physics and Institute of Physics and Applied Physics,
Yonsei University, Seoul 120-749 Korea}

\date{\today}

\maketitle

\begin{abstract}
We study the analytic structure of the S-matrix which is obtained
from the reduced Wheeler-DeWitt wave function describing
spherically symmetric gravitational collapse of massless scalar
fields. The complex simple poles in the S-matrix lead to the wave
functions that satisfy the same boundary condition as quasi-normal
modes of a black hole, and correspond to the bounded states of the
Euclidean Wheeler-DeWitt equation. These wave function are
interpreted as quantum instantons.
\end{abstract}
\pacs{04.70.Dy, 04.60.Kz}

In the previous work \cite{bak1} we studied quantum mechanically
the self-similar black hole formation by collapsing scalar fields
and found the wave functions that give the correct semi-classical
limit. The reduced Wheeler-DeWitt equation for gravity belongs to
an exactly solvable Calogero type system with an inverted
potential whose attractive inverse square and repulsive square
potential terms give rise to a potential barrier. The boundary
condition for black hole formation was that the wave function has
both the incoming and the outgoing flux at spatial infinity and
only the incoming flux toward the black hole singularity.

Of particular interest is the subcritical case, in which a black
hole can be formed through quantum tunneling. Due to the time
reversal symmetry, however, the subcritical wave function may be
given an interpretation of the reversal process of black hole
formation, that is, the decay of the black hole \cite{tomimatsu}.
Then the wave function for black hole decay should have a purely
outgoing flux. This wave function is somehow reminiscent of the
gravitational wave from a perturbation of a black hole
\cite{regge}. Moreover, for a certain discrete spectrum of complex
frequencies there occur quasi-normal modes that have both purely
outgoing modes at spatial infinity and purely incoming ones at the
horizon of the black hole \cite{vish}.

In this paper we study the pole structure of the S-matrix which is
obtained from the wave function for black hole formation. The
boundary condition that the wave function should have a purely
outgoing flux at spatial infinity and a purely incoming one at the
classical apparent horizon leads to a discrete spectrum of complex
parameters $(c_0)$. It is further shown that this wave function
can be obtained through the analytical continuation of a bounded
state of the corresponding Euclidean Wheeler-DeWitt equation. Just
as quasi-normal modes of perturbations of a black hole can be
interpreted as instantons \cite{quasi}, these exact wave functions
of the quantum theory for black hole decay may be interpreted as
quantum instantons

The spherically symmetric geometry minimally coupled to a massless
scalar field is described by the reduced action in
$(1+1)$-dimensional spacetime of which the Hilbert-Einstein action
is
\begin{equation}
S = \frac{1}{16\pi} \int_M d^4x \sqrt{-g}~ \left[ R - 2 \left(\nabla \phi
\right) ^2 \right] + \frac{1}{8\pi} \int_{\partial M} d^3 x K \sqrt{h}.
\end{equation}
The reduced action is
\begin{equation}
S_{sph} = \frac{1}{4} \int d^2 x \sqrt{-\gamma} ~r^2 \left[ \left\{^{(2)}
R(\gamma) + \frac{2}{r^2} \left(\left(\nabla r \right)^2 + 1\right)
\right\} -2 \left(\nabla\phi\right)^2 \right],
\end{equation}
where $\gamma_{ab}$ is the $(1+1)$-dimensional metric.  The spherical
spacetime
metric is
\begin{equation}
ds^2 = -2 du~dv + r^2 d\Omega_2^2 ,
\end{equation}
where $d\Omega_2^2$ is the usual spherical part of the metric, and $u$
and $v$ are null coordinates.  The self-similarity condition is imposed
such that
\begin{equation}
r = \sqrt{-uv}~ y(z), \quad \phi = \phi(z),
\end{equation}
where $z = +v/(-u) = e^{-2\tau}$, $y$ and $\phi$ depend only on
$z$. We introduce another coordinates $(\omega, \tau)$
\begin{equation}
u = -\omega e^{-\tau}, \hspace{3mm} v = \omega e^\tau ,
\end{equation}
to rewrite the metric as
\begin{equation}
ds^2 = -2 N^2 (\tau) \omega^2 d\tau^2 + 2d\omega^2 + \omega^2 y^2
d\Omega_2^2,
\end{equation}
where $N(\tau)$ is the lapse function of the ADM formulation.

The classical solutions of the field equations were obtained by
Roberts \cite{Roberts}, and studied in connection with
gravitational collapse by others \cite{others}.  Classically black
hole formation is only allowed in the supercritical case ($c_0
>1$), but even in the subcritical situation there are quantum
mechanical tunneling processes to form a black hole of which the
probability is semiclassically calculated \cite{bak2,bak1}.

In our previous work \cite{bak1} we quantized the system
canonically with the ADM formulation to obtain the Wheeler-DeWitt
equation for the quantum black hole formation
\begin{equation}
\left[\frac{\hbar^2}{2K} \frac{\partial^2}{\partial y^2} -
\frac{\hbar m_P^2}{2Ky^2} \frac{\partial^2}{\partial\phi^2} -
K\left(1-\frac{y^2}{2}\right) \right] \Psi(y,\phi) = 0, \label{wd
eq}
\end{equation}
where $K/\hbar \equiv (m_p^2/\hbar^2)(\omega_c^2/2)$ plays the
role of a cut-off parameter of the model, and we use a unit system
$c=1$.  The wave function can be factorized to the scalar and
gravitational parts,
\begin{equation}
\Psi(y,\phi) = \exp\left(\pm i \frac{K c_0}{\hbar^{1/2} m_P} \phi
\right) \psi(y).
\end{equation}
Here the scalar field part is chosen to yield the classical
momentum $\pi_\phi = \hbar K y^2 \dot{\phi}/m_P^2 N = \pm Kc_0$,
where $c_0$ is the dimensionless parameter determining the
supercritical ($c_0
> 1$), the critical ($c_0 =1$), and the subcritical ($1>c_0 >0$)
collapse.

Now the gravitational field equation of the Wheeler-DeWitt
equation takes the form of a Schr\"{o}dinger equation
\begin{equation}
\left[\frac{-\hbar^2}{2K} \frac{d^2}{dy^2} + \frac{K}{2}\left(2 -
y^2 - \frac{c_0^2}{y^2} \right) \right] \psi(y) =0. \label{grav
eq}
\end{equation}
The solution describing black hole formation was obtained in Ref.
\cite{bak1}:
\begin{equation}
\psi_{BH} (y) = \left[\exp \left( \frac{-i}{2} \frac{K}{\hbar} y^2
\right) \right] \left(\frac{K}{\hbar} y^2 \right)^{\mu} M\left( a,
b, i\frac{K}{\hbar}y^2\right), \label{bh wave}
\end{equation}
where $M$ is the confluent hypergeometric function and
\begin{equation}
a = \frac{1}{2} - \frac{i}{2\hbar} (Q + K), \quad b = 1 -
\frac{i}{\hbar}Q, \quad \mu = \frac{1}{4} - \frac{i}{2\hbar}Q
\end{equation}
with
\begin{equation}
Q = \left( K^2 c_0^2 -\frac{\hbar^2}{4}\right)^{1/2}.
\end{equation}
Using the asymptotic form \cite{asymptotic} at spatial infinity
\begin{equation}
\psi_{BH} (y) \simeq   \frac{\Gamma (b)}{\Gamma(b - a)} e^{i\pi a}
\left(i\frac{K}{\hbar} y^2 \right)^{\mu - a} e^{-(i/2) (K/\hbar)
y^2} + \frac{\Gamma (b)}{\Gamma(a)} \left(i\frac{K}{\hbar} y^2
\right)^{\mu + a - b} e^{(i/\hbar)(K/\hbar) y^2}, \label{asymp}
\end{equation}
we obtain the S-matrix component describing the reflection rate
\begin{equation}
S = \frac{\Gamma (b - a)}{\Gamma(a)} \frac{(iK/\hbar)^{2 a -
b}}{e^{i\pi a}}.
\end{equation}
From the S-matrix follows the transmission rate for black hole
formation
\begin{eqnarray}
\label{trans} \frac{j_{trans}}{j_{in}} & = & 1 - |S|^2 \nonumber
\\ & = & 1- \frac{\cosh \frac{\pi}{2\hbar}(Q+K)}{\cosh
\frac{\pi}{2\hbar}(Q-K)} e^{-(\pi/\hbar) Q} ,
\end{eqnarray}
where $\left|\Gamma \left(\frac{1}{2}+ix\right)\right|^2 =
\frac{\pi}{\cosh(\pi x)}$ is used.  Equation (\ref{trans}) gives
the probability of black hole formation for the supercritical,
critical, and subcritical $c_0$-values.

We now consider the analytic structure of the S-matrix: it is an
analytic function of $Q$ and $K$ with simple poles which can be
explicitly shown as
\begin{equation}
\label{Smatrix} S = \sum_{N=0}^{\infty} \frac{1}{(Q - K)/\hbar
+i(2N+1)} \left( \frac{2i e^{-(\pi/2\hbar) K - i(K/\hbar) \ln
(K/\hbar)}}{N! \Gamma(- N - i (K/\hbar))}\right). \label{pole1}
\end{equation}
The poles reside in the unphysical region of the parameter space of $Q$
and $K$:
\begin{equation}
\label{Q} Q - K = - i \hbar (2N+1) , \quad (N = 0,1,2, \cdots).
\label{17}
\end{equation}
It should be remarked that these poles make the first term of Eq.
(\ref{asymp}) vanish since
\begin{equation}
b - a = - \frac{i}{2\hbar}(Q -K) + \frac{1}{2} = - N, \quad (N =
0, 1, 2, \cdots). \label{pole2}
\end{equation}
The second term of Eq. (\ref{asymp}) has a purely outgoing flux at
spatial infinity. The wave function near the apparent horizon,
which can be obtained by the steepest descent method in the
Appendix of Ref. \cite{bak1} and by taking the large
$(K/\hbar)$-limit, leads to the flux
\begin{equation}
j_{AH} \simeq A^2 (y) \Biggl\{\frac{1}{2} y (1 - y^2)
\Biggl[\frac{(y^4 + c^{*2} - 2y^2)^{1/2} + (y^4 + c^2 -
2y^2)^{1/2}}{\Bigl((y^4 + c^{*2} - 2y^2) (y^4 + c^2 - 2y^2)
\Bigr)^{1/2}} \Biggr] - \frac{1}{2} (c_0^* + c_0) \frac{1}{y}
\Biggr\},
\end{equation}
where $A(y)$ denotes an amplitude, a real function, and $c_0 = 1 -
i(\hbar/K)(2N+1)$ from Eq. (\ref{17}) and $c = (c_0^2 -
\hbar^2/4K^2)^{1/2} - i (\hbar/K)$. At the apparent horizon
$y_{AH} = c_0/\sqrt{2}$, the wave function has an incoming flux.
Therefore, the poles are the outcome of the same boundary
condition used to find quasi-normal modes of a black hole
\cite{vish}. Note that the wave function (\ref{bh wave}) has also
the purely incoming flux toward the black hole singularity at $y =
0$.

A few comments are in order. First, for physical processes of
gravitational collapse there can not be poles because $K$ and
$c_0$ are real-valued. In ordinary quantum mechanics, the poles of
S-matrix occur at the bound states \cite{poles}, and in
relativistic scattering at the resonances or the Regge poles
\cite{Regge}. Our case is analogous to a meta-stable quantum
mechanical system of which poles are identified with
quasi-stationary states that describe the decay of a particle
through a potential barrier. Second, we calculated quantum decay
rate of a black hole as a reversed process of gravitational
collapse through a barrier by quantum tunneling. This quantum
decay process, first studied by Tomimatsu \cite{tomimatsu}, is a
distinctively different decay channel from the Hawking radiation
process. It will be interesting to investigate both processes
present in one model. Third, it should be pointed out that our
discussion based upon the similarity of the boundary condition on
the wave function with quasi-normal modes seems to have no deeper
physical connection more than analogy because our model works only
for a dynamical stage of gravitational collapse and its reversed
process, rather than the quasi-stationary stage at late times.

Recalling that the poles in Eqs. (\ref{pole1}) or (\ref{pole2})
result from the potential barrier and that the exponential
behavior of the Wheeler-DeWitt equation under a potential barrier
describes a Euclidean geometry, we turn to the Euclidean theory of
gravitational collapse. In the Euclidean theory the Wheeler-DeWitt
equation has oscillatory wave functions and a well-defined
semiclassical limit even under the potential barrier of the
Lorentzian theory \cite{kim}. The Euclidean geometry with the
metric
\begin{equation}
ds_E^2 = 2 N^2 (\tau) \omega^2 d\tau_E^2 + 2d\omega^2 + \omega^2
y^2 d\Omega_2^2,
\end{equation}
leads to the Wheeler-DeWitt equation
\begin{equation}
\left[- \frac{\hbar^2}{2K} \frac{\partial^2}{\partial y^2} +
\frac{\hbar m_P^2}{2Ky^2} \frac{\partial^2}{\partial\phi^2} -
K\left(1-\frac{y^2}{2}\right) \right] \Psi_E (y,\phi) = 0.
\end{equation}

According to the transformation rule  $i \pi_{\phi}
\leftrightarrow \pi_{E, \phi}$ of the scalar field momenta between
the Lorentzian and Euclidean geometries \cite{kim}, the wave
function has the form
\begin{equation}
\Psi_E (y, \phi) = \exp\left(\mp \frac{K c_0}{\hbar^{1/2} m_P}
\phi\right) \psi_E (y).
\end{equation}
The Wheeler-DeWitt equation reduces to the gravitational field
equation
\begin{equation}
\left[\frac{-\hbar^2}{2K} \frac{d^2}{dy^2} + \frac{K}{2}\left( y^2
+ \frac{c_0^2}{y^2} -2 \right) \right] \psi_E (y) =0.
\label{e-grav eq}
\end{equation}
Notice that this is a variant of Calogero models with the
Calogero-Moser Hamiltonian \cite{Calogero}, but the energy
eigenvalue is fixed, and only a quantized $c_0$ is allowed. Since
Eq. (\ref{e-grav eq}) can also be obtained from the Lorentzian
equation (\ref{grav eq}) by letting
\begin{equation}
K = i K_E,
\end{equation}
one may obtain, through the analytical continuation of Eq.
(\ref{bh wave}), the solution to Eq. (\ref{e-grav eq})
\begin{equation}
\psi_E (y) = \left[\exp \left( \frac{1}{2} \frac{K_E}{\hbar} y^2
\right) \right] \left(\frac{K_E}{\hbar} y^2 \right)^{\mu_E}
M\left( a_{E}, b_{E}, - \frac{K_E}{\hbar} y^2\right),
\label{e-wave}
\end{equation}
where
\begin{equation}
a_{E} = \frac{1}{2} + \frac{1}{2\hbar} (Q_E + K_E), \quad b_{E} =
1 + \frac{Q_E}{\hbar}, \quad \mu_E = \frac{1}{4} +
\frac{1}{2\hbar}Q_E
\end{equation}
with
\begin{equation}
Q_E =\left( K_E^2 c_0^2 + \frac{\hbar^2}{4}\right)^{1/2}.
\end{equation}
The asymptotic form of Eq. (\ref{e-wave}) leads to the bounded
states only when
\begin{equation}
b_{E} - a_{E} = - N, \quad (N = 0, 1, 2, \cdots),
\end{equation}
that is, the condition is satisfied
\begin{equation}
Q_E - K_E = - \hbar (2N +1). \label{bound}
\end{equation}
The condition (\ref{bound}) is identical to the pole position of
the S-matrix with $K=iK_E$ given in Eqs. (\ref{Q}) and
(\ref{pole2}).

A few remarks are in order. First, the quantum solution
(\ref{e-wave}) is analogous to an instanton in the sense it is a
solution in the Euclidean sector, but is not in the strict sense
because the Wheeler-DeWitt equation is already a quantum equation,
not a classical one. The semiclassical result from the
Bohr-Sommerfeld quantization rule
\begin{equation}
\frac{\pi}{2} \frac{K_E}{\hbar} (1 - c_0 ) = \pi \Biggl(N +
\frac{1}{2} \Biggr), \quad (N = 0, 1, 2, \cdots), \label{inst}
\end{equation}
is the large $(K_E/\hbar)$-limit of the exact result
(\ref{bound}). The instantons, the left hand side of Eq.
(\ref{inst}), provides semiclassically the probability of
tunneling process \cite{bak2}. Second, the correspondence between
the poles and the Euclidean polynomial solutions breaks down for
large $N$.  While the poles contribute for all $N$ without limit,
the normalizable Euclidean solutions exist only for $N <
K_E/2\hbar$. The polynomial solutions for large $N$ are well
defined, but are not normalizable.  We have not yet understood
these nonnormalizable solutions. Finally, we consider the
classical field equations corresponding to the poles of the
S-matrix. In the Lorentzian geometry the relevant equations are
\begin{equation}
\frac{d\phi}{d\tau} = \frac{c_0}{y^2},
\end{equation}
\begin{equation}
\left( \frac{dy}{d\tau} \right)^2 = K^2 \left( -2 + y^2 +
\frac{c_0^2}{y^2} \right),
\end{equation}
where $c_0 \simeq 1- i \hbar (2N+1) / K$, for large $K$. The
complex $c_0$ implies complex $\frac{d\phi}{d\tau}$ and
$\frac{dy}{d\tau}$, which may be imagined as a bound state like
complex momentum in quantum mechanics and requires a complex
spacetime metric. In the Euclidean geometry ($K = iK_E$) these
classical equations are the same as those equations with quantized
$c_0$ in the tunneling region in Ref. \cite{bak2}.

\begin{acknowledgements}
We would like to express our appreciation for the warm hospitality
of CTP of Seoul National University where this paper was
completed. K.S.S. and J.H.Y. were supported in part by BK21
Project of Ministry of Education. D.B. was supported in part by
KOSEF-98-07-02-07-01-5, S.P.K. by KOSEF-1999-2-112-003-5, S.K.K.
by KRF-99-015-DI0021, and J.H.Y. by KOSEF-98-07-02-02-01-3.
\end{acknowledgements}

\end{document}